\documentstyle[12pt]{article}
\begin{document}

\begin{center}
{\bf SEARCHING FOR COLOR COHERENT EFFECTS\\
 AT INTERMEDIATE {\boldmath $Q^2$}\\
VIA DOUBLE SCATTERING PROCESSES\\}
\vspace{2mm}
K.~Egiyan$^1$, L.~Frankfurt$^{2,3}$, W.~R.~Greenberg$^{4,5}$, G.~A.~Miller$^4$,
\\M.~Sargsyan$^{1,2}$ and M.~Strikman$^{3,6}$\\

\vspace{1cm}
\begin{center}
{\it $^1$ Yerevan Physics Institute, Armenia\\
$^2$ School of Physics and Astronomy, Tel Aviv University, Israel\\
$^3$ St. Petersburg Nuclear Physics Institute, Russia\\
$^4$ University of Washington, Seattle, WA, USA\\
$^5$ Present Address: University of Pennsylvania, Philadelphia, PA, USA\\
$^6$ Pennsylvania State University, University Park, PA, USA}
\end{center}

\vspace{2mm}
{\bf ABSTRACT}
\end{center}
\vskip 1mm
We propose
that  measuring
the $Q^2$ dependence of
the number of final-state
interactions   of the recoil protons
in quasi-elastic electron
scattering from  light nuclei is a new method to investigate Color
Coherent effects at {\bf intermediate} values of
$Q^2$ ({$\sim$ few $(GeV/c)^2$}).
This is
instead of measuring events without final-state
interactions. Our calculations indicate
that  such
measurements
 could reveal significant color
transparency  effects for
the highest of the energies initially available at CEBAF.
Measurements that detect more than one hadron
in the final state, which require
the use of large acceptance ($4\pi$) detectors, are required.

\vspace{10mm}
\section{Introduction}

How does quantum chromodynamics QCD work?
Although the purely perturbative regime is well understood, much more
needs to be done to answer the question.
To be specific, consider  the process of
elastic electron-nucleon scattering. At high enough values of the momentum
transfer Q$^2$, pQCD is valid and
the dominant contributions to the form factor arise from components in
which the quarks and gluons are closely separated~\cite{bpqcd}.
Such components have been called point-like configurations PLC~\cite{fs,fs85}.
However, it is not at all clear that this regime is relevant
for the Q$^2$ for which experiments exist~\cite{isgur,ar}.

Another idea~\cite{fms92} is that PLC may arise from
non-perturbative effects.  In this case
the PLC can be thought of as precursors to the dominance of pQCD.
Indeed, the work of Ref.~\cite{fms92}
shows that
realistic quark models of a nucleon which contain a
Coulomb type interaction at small interquark distances and Skyrmion models
both allow a PLC to form
starting
from a momentum transfer as small as $1-2 \, (GeV/c)^2$.
The opposite behavior is expected in mean-field quark-models of a
nucleon, and in Chiral Lagrangian
models  where a nucleon is considered
as a structureless particle with a meson cloud.

Thus, the pressing problem now is to find experimental measurements that
help to distinguish between the two classes of models.
The ideas of pQCD lead to the suggestion~\cite{bct,mct} that
the A-dependence of quasi-exclusive processes
\begin{equation}
l(h) + A \rightarrow l(h) + p + (A-1)
\end{equation}
could be used to determine the configurations that dominate in hard two-body
reactions. If PLC are produced,
QCD color screening causes the ejectile-nucleon interaction to be small, and
the ejected object
would escape from the nucleus with no or reduced final-state interactions.
Hence
the cross section of reaction (1) would be
proportional to A. This is the  Color Transparency (CT) phenomenon.
This prediction of
a spectacular change of the A-dependence of the cross section of reaction (1)
has led to a number of  further theoretical analyses~
\cite{FLFS,jm,FSZ,rp,JK93,BBK93}
 and to  the first
attempts  to observe the  phenomenon using the BNL
proton beam~\cite{BNL} and
ongoing experimental investigations at SLAC (NE18)~\cite{NE18} and
BNL~\cite{EVA}.

But there is an impressive
practical problem which emerges in looking for CT effects in actual
experiments. The PLC expands rapidly to the size of a normal
hadron while propagating through the nucleus, unless the Lorentz
factor is very large~\cite{FLFS,jm}. This does not occur for intermediate
values of Q$^2$.
Therefore, the observation of CT effects
at intermediate values of Q$^2$ requires the suppression of
wavepacket expansion effects.  This can be achieved by using the
lightest nuclear targets, where the propagation distances  are small.
  In
this case, the transparency is close to unity, so the effects of CT in
$(e, e'p)$ reactions can not be large.
  However, if we select a process where the produced
system interacts in the final state, a double scattering event, then the
color coherent effects would manifest themselves as a decrease of the
probability for final-state interactions with increasing $Q^2$.
Then one would observe a $(e,e^\prime pp)$ reaction with one proton having
a momentum close to that of the virtual photon and a second proton
other with a
momentum $p_2$ of about 400~$MeV/c$, large enough so that its production
is dominated by the effects of
final-state interactions.
The
obvious advantage of looking for the processes with rescattering is that in
this case it is possible to observe an effect decreasing from the value
expected without CT (eikonal approximation)
to zero.  Thus, the measured cross
section is to be compared with a vanishing quantity so that the relevant
ratio of cross sections runs from 1 to infinity.

The value of $p_2\approx $400 MeV/c is chosen as large enough
so that its origin is the effects of nuclear correlations.
This is also
 but small enough (-t=0.15 GeV$^2$)
so that the
forward peaked rescattering is allowed to occur.
The effects of correlations are small,  but may can not be entirely
negligible.
Indeed,
if it happens that the  final state interactions are removed by the effects of
color transparency, a non-zero
cross section could be caused by the correlation effects.
Moreover, the kinematics of scattering from a pair of nucleons causes
the angular distributions for an
 correlation dominated e,e'pp reaction to be substantially different from
rescattering dominated ones.
 Thus one could eliminate
this unwanted ``background"
using this signature and
the  4 $\pi$ detector.

We consider two models  to
explore this new possibility to observe CT effects.
The first calculation uses the
quantum diffusion model of~\cite{FLFS} and a Monte Carlo simulation. This
is an intuitive calculation that allows us
to consider a wide range of nuclei and to account also for the
dependence of the rescattering amplitude
slope on the size of a rescattered configuration. The second calculation
employs
the three-state model of Ref.~\cite{FGMS}. This
 allows us to perform a quantum
mechanical calculation for the $^3He$ target,
 though with an oversimplified wave
function.
Both models are schematic and do not allow us to reliably estimate the
effects of the background of (e,e'pp ) events produced by nuclear
correlations.

 It is worth emphasizing
that in both models the  CT  effect predicted for the $A(e,e^\prime p)$
process is rather small and does not contradict preliminary data of the
NE-18 experiment. We  demonstrate that both models lead to quite
substantial  CT effects for the rescattering reaction of interest.

Our results emphasize the need for both
experimental study of this reaction,
feasible at CEBAF, and for more refined calculations including realistic
wave functions of $^{3,4}He$ and short range correlations for heavy
nuclei.

\section{Probabilistic estimation of rescattering in $(e,e'pp)$}

In general, the
calculation of $A(e,e'NN)$ processes is rather cumbersome. So
we
will use a simple probabilistic description  to
illustrate the importance of color screening (reduced interactions of small
objects)
and the space-time picture of a PLC evolving to a normal state. A
quark-basis description is used to describe this evolution.
This intuitive approach
gives us the possibility of
considering  heavier targets than for the next section.
Thus, in this approximation,  the  $(e,e'pp)$ cross section
can  be represented as:
\begin{eqnarray}
{d^9\sigma \over dE_{e'} d\Omega_{e'} d^{3}p_{f'}
d^{3}p_{r} }  &=&   \int d\Omega_{f}
{d\sigma^{PWIA}_{qe} \over dE_{e'} d\Omega_{e'} d\Omega_{f} } \cdot
\Phi^{2}_{A-1}(|\vec p_{r0}|)\cdot {2|\vec p_{f}| \over E_{f'}}
\cdot \nonumber \\
&& \nonumber \\
&&  \langle \int \rho(l){d\sigma^{pp}(l,Q^{2})\over dt}dl
\exp \left( -\int_{z}^{\infty} \sigma_{tot}(l',Q^{2})\rho(l')dl'
\right) \rangle \cdot \nonumber \\
&& \nonumber \\
&& \delta(q_{0}+M_{A}-E_{f'}-E_{r}-E_{A-2})
\label{eq:cs}
\end{eqnarray}
where $(E_{f'},\vec p_{f'})$ and $(E_{r},\vec p_{r})$ are the energy and
momenta of the produced (by virtual photon with energy and momentum
($q_{0},\vec q$)) and
rescattered (by $pp$-elastic scattering) protons respectively,
${\vec p_{r0}=\vec p_{r}-(\vec p_{f}-\vec p_{f'})}$, where $\vec p_{f}$ is
the momentum of produced protons before the rescattering.
For the simple case of quasifree rescattering
${E_{A-2}\approx M_{A-2}+p_{A-2}^{2}/2M_{A-2}}$,
where ${\vec p_{A-2}=\vec q - \vec p_{f'} - \vec p_{r}}$ (in
calculations we specify the kinematics where ${\vec p_{A-2} \approx 0}$).
The ${d\sigma^{PWIA}_{qe} \over dE_{e'} d\Omega_{e'} d\Omega_{f} }$
represents the quasielastic $(e,e'p)$ cross section in the plane wave
impulse approximation (PWIA), $\Phi^{2}_{A-1}(|\vec p_{r0}|)$ is the
momentum distribution of rescattered protons in $A-1$ residual nucleus.
The term in the ``$\langle \  \rangle$" gives the probability for
the processes where final nucleon experience only one elastic collision.
The  brackets  means that the integral over the
transverse coordinates is taken. (Those coordinates are left implicit in
Eq.~(\ref{eq:cs}) to simplify the notation.)
The quantity ${d\sigma^{pp}(l,Q^{2})\over dt}$ represents the elastic
scattering between a proton and a
PLC that has moved a distance $l$ from its point
of formation.

Eq.~(\ref{eq:cs}) accounts for the geometry of the process, but not
the quantum mechanical averaging over the nuclear
 wavefunction. This may introduce
uncontrollable errors in our calculations, especially in the case of heavy
nuclei where interference effects between
elastic rescattering and nucleon
correlations could be large. However, the present calculation demonstrates
the
significant size of color transparency effects
and the noticeable dependence on momentum transfer.

It is easy to check that after integrating the Eq.~(\ref{eq:cs}) by
$p_{f'}$ and $p_{r}$ and summing over  elastic rescattering states one recovers
the conventional  probabilistic description
for the quasielastic $(e,e'p)$ cross section. Note that above equation is
better suited for the case of quasifree (e,e') scattering at $x = {Q^{2}\over
2mq_{0}} \approx 1$, where Fermi motion effects are a correction.

To estimate the color coherent effects in Eq.~(\ref{eq:cs}) we introduce the
differential cross section for the PLC elastic scattering with momentum
transfer $t$ and at the distance $l$ from the point where the photon has been
absorbed:
\begin{equation}
{d\sigma^{pp}(l,Q^{2})\over dt} = {\sigma^{2}_{tot}(l,Q^{2}) \over 16\pi}
\cdot e^{bt}\cdot {G_{N}^{2}(t\cdot\sigma_{tot}(l,Q^{2})/\sigma_{tot})
\over G_{N}^{2}(t) }
\label{eq:sigma-pp}
\end{equation}
Here $\sigma_{tot}$ is a proton-nucleon total cross section, $b$ is the
slope of elastic $NN$ amplitude, $G_{N}(t)$ ($\approx (1-t/0.71)^{2}$) is the
Sachs form factor. The last factor in Eq.~(\ref{eq:sigma-pp}) accounts for
the difference between elastic scattering for point-like and average
configurations, which is based on the observation that $t$ dependence of
${d\sigma^{h+N\rightarrow h+N}/dt
\sim~G_{h}^{2}(t)\cdot G_{N}^{2}(t)}$.

In Eq.~(\ref{eq:sigma-pp}) $\sigma_{tot}(l,Q^{2})$ is the effective total
cross section of the interaction of the PLC at the distance $l$ from
the interaction point. The quantum
diffusion model~\cite{FLFS} provides the estimate:

\begin{equation}
\sigma _{tot}(l,Q^{2}) = \sigma_{tot} \left \{ \left ( {l \over l_{h} } +
{ \langle r_{t}(Q^2)^{2} \rangle  \over \langle r_t^{2} \rangle }
(1-{l \over l_{h}}) \right )  \Theta (l_{h}-l) +
\Theta (l-l_{h}) \right\}
\label{eq:sigm-ct}
\end{equation}
where  ${l_h = 2p_{f}/\Delta~M^{2}}$,
with  ${\Delta~M^{2}=0.7~GeV^{2}}$.
Here ${\langle r_{t}(Q^2)^{2} \rangle}$ the average transverse size squared
of the configuration produced at the
interaction point. In several realistic models considered in~\cite{fms92}
it can be approximated as
${ {\langle r_{t}(Q^2)^2 \rangle \over \langle r_t^2\rangle }
\sim{1\, GeV^2\over Q^2}}$ for $Q^2~\geq~1.5~GeV^2$.
Note that the effects of expansion cause the results to
rather insensitive to the value of
this ratio whenever it is much less than unity.

It is tedious, but  not difficult,  to demonstrate that the semiclassical
calculation which includes quantum diffusion gives the exact result in QCD for
$\sigma_{tot}(l,Q^2)$ at
sufficiently large $Q^2$ in the leading logarithmic approximation
and beyond. At moderate $Q^2$, quantum diffusion is a guess based on an
analogy with pQCD and on the success of related approaches in the
description of final states in $e^+~e^-~\rightarrow~\rm hadrons$.
The success of dispersion sum rules in describing semi-hard processes
indicates that  matrix elements of correlators of currents can be effectively
be calculated in either of the quark gluon or hadron bases.

Note that Eq. (\ref{eq:cs}) has been obtained by using
classical mechanics to calculate the
rescattering processes.  The difference between this and the
semiclassical
approximation (see e.g.~\cite{FLFS}) is that
in the classical mechanical description one averages the absolute square of
the transition matrix element instead of averaging the amplitude and then
squaring.
The
shortcoming of this approach to rescattering processes is that it
gives no possibility  to account properly for the
correlations between nucleons.
Thus, we use a simple one-body Wood-Saxon
parametrization of the
nuclear density function $\rho(r)$ appearing in Eq.~(\ref{eq:cs}).
The authors hope to improve this in future publications.

We turn to the results.
 First, Fig.~1(a) and Fig.~2(a)  show the  predictions of
probabilistic approach and independent particle model for
the standard transparency T defined as
${T={\sigma(e,e'p)^{exp} \over \sigma(e,e'p)^{pwia}}}$
{}~(see~e.g. Refs.~\cite{NE18,FSZ93,ob}). We find
that predictions of this simplified approach  agree well with more
conventional quantum mechanical optical model and CT
calculations ~\cite{FSZ,ob} within a few percent.
The $(e,e'pp)$ calculations presented in Figs.~1(b-d),~2(b-d) are  performed
for
the  kinematics where ${\vec p_{f} \sim \vec p_{f'} \sim \vec q}$ and
${|\vec q|~\gg |\vec p_{r}|~\approx~0.4~GeV/c}$  to suppress Fermi motion
and
evaporation effects.  For  $pp$ rescattering  this corresponds to
the transferred momentum ${t~\approx~-0.15~GeV^2}$.
In Fig.~1(b) and Fig.~2(b) we present results of calculations of
the transparencies defined as:
\begin{equation}
T^{el} = \left( {d\sigma^{9} \over dE_{e'} d\Omega_{e'} d^{3}p_{f'}
d^{3}p_{r} }\right)  / \left ( {d\sigma^{5}
\over dE_{e'} d\Omega_{e'} d\Omega_{f} } \right )^{PWIA}.
\label{eq:T-el}
\end{equation}
These are computed in Glauber approximation (solid curve) and including the
effects of CT. The solid and dashed curves have a ratio of about 1.5
for Q$^2\approx 7$GeV$^2$/c$^2$ so that the
effects of CT
are substantial.

Another useful quantity is the ratio of the cross sections of double scattering
$(e,e'p_{f}p_{r})$ and $(e,e'p_{f})$ processes studied above. Note that
the uncertainties in the estimate of PLC production  cancel to large extent
in this ratio. Another advantage is that there  is more
sensitivity to CT effects. Indeed  using the  T  and the $T^{el}$
defined above (see also~\cite{cebafprepr}) one gets:

\begin{equation}
{T_{el}\over T} = {\sigma(e,e'pp)^{exp} \over \sigma(e,e'p)^{exp}}
\simeq { \langle \int \rho(l){d\sigma^{pp}(l,Q^{2})\over dt}dl
\exp \left( -\int_{z}^{\infty} \sigma_{tot}(l',Q^{2})\rho(l')dl' \right)
\rangle \over
\langle \exp \left( -\int \sigma_{tot}(l,Q^{2})\rho(l)dl \right)\rangle },
\label{eq:T_r}
\end{equation}
where the effects of CT cause the numerator to decrease and the denominator
to increasing. As a result the
 difference between CT and Glauber calculations is more
pronounced for these quantities - compare the solid and dashed curves in
 Figs.~1(c),~2(c). Note that from the
experimental point of view, Eq.~(\ref{eq:T_r}) represents the
ratios of quantities measured in the same run of experiment, that allow to
avoid  many problems (such as necessity to
calculate a denominator in $T$, the radiative corrections, etc.).

This point is further demonstrated in the Fig.~1(d) and Fig.~2(d) where
calculations of CT for $T, T_{el},
T_{el}/T$  are normalized by the corresponding quantities computed
in the Glauber approximation. The effect of deviation from the
 Glauber calculation is the
smallest for T - solid curves, larger for $T_{el}$ -dashed curves, and the
largest for the $T_{el}/T$ ratios - dotted curves.
 These figures demonstrate that  color transparency
is a
significant effect for the case of double rescattering processes   even at
intermediate $Q^2$. Note that results of calculation in this section are
consistent with the predictions of three state model considered in the
next section.

The results displayed above
are for double rescattering processes at fixed $t$ or at
fixed relative transverse ( to the direction of virtual photon) momenta of
two final proton.
These cross sections should tend to zero when
the color transparency phenomena become important. Another
observable is the $t$ ($p_t$) dependence of $(e,e'2p)$ cross section at a
fixed (large) value of $Q^2$. Here, one expects
an {\bf enhancement} of the $(e,e'pp)$
cross section starting at some transfered momenta $t$ for
the CT cases compared to that of the Glauber approximation~\cite{fs}. This
is because at sufficiently large values of $t$ the elastic
rescattering are dominated by the interaction
of the nucleon PLC components.
Here the underlying physics is that
large angle scattering occurs
anyway in a PLC, and if the  initial state
is in PLC  one does not  have to pay the
price of finding the projectile in
PLC.

The t-dependence of the cross section ratios is examined in Fig.~3.
We use the parametrization of $pp$
cross sections  in Eq.~(\ref{eq:sigma-pp}). The dashed curve shown in this
figure indicates a significant
enhancement of $\sigma^{CT}/\sigma^{Glauber}$
${|t| \geq 1~GeV^2}$ for
 sufficiently high ${Q^2~(\geq 10~GeV^2)}$. The solid curve
for Q$^2$=4 GeV$^2$ shows a slower t-dependence. Furthermore, there is not
much of a change as -t is changed from 0 to 0.16 GeV$^2$.
This supports the
approximations  made in Sec.~3, where this dependence is
neglected.

\section{Calculation of  $^3He(e,e'pp)$ in the three state model}

The measurement
of the $^3He(e,e'pp)$ reaction of our interest involves
detecting a high momentum proton carrying almost
all of the virtual photon's three-momentum, and another
of moderate momentum, about 400~$MeV/c$.  The value 400~$MeV/c $
is chosen as small compared with virtual photon's
momentum and large compared with the momentum of a bound proton.
If such a 400 MeV/c proton is
detected, we can be almost certain that it was produced as a result of
a final state interaction. If CT occurs, there are no final state
interactions and our $^3He(e,e'pp)$ reaction will not take place.

In order to
make an illustrative calculation for
such a process, we use the three-state model of Ref.~\cite{FGMS}.
In this model, the resonance and continuum regions excited in diffractive
dissociation processes are each modelled as one state.
Since the rescattering process involves only
a small momentum transfer,
and because
the major contribution at moderate energies is given by the color
fluctuations near the average size  we shall
ignore the variation of the t dependence of the scattering
amplitude
with the  distance from  the point of hard the interaction.
This approximation means that we
neglect any experimentally observed  difference between the
form factors of different  states.
The possible effect of neglecting this t-dependence is examined
in section 2, where the effects are found to be small.

The matrix element for our (e,e',pp) process is given by
\begin{equation}
{\cal M}_1(\vec p_1, \vec p_2, \vec p_3)
= \langle N_1, {\vec p}_1; N_2, {\vec p}_2; N_3, {\vec p}_3 |
\sum_{i\neq j}
T_j G_0 T_H^{(i)} | ^3He \rangle,
\label{eq:heme}
\end{equation}
where the hard scattering operator $T_H^{(i)}$
acts on the $i$'th proton, which
propagates a wavepacket, via the free Green's function G$_0$, until a
soft final-state interaction T$_j$ leads to the knockout of the j'th
proton. Most of our interest here concerns the
(e,e'pp) reaction, but we also compute cross sections for the
(e,e'N$^*$p) and (e,e'N$^{**}$p) processes.
 The expression for N$^*$ or N$^{**}$ production is similar to that
of Eq. (\ref{eq:heme}) except that the $N_1,\vec p_1$
is replaced by $m_1,\vec p_{1m}$ where $m$ is either of the two resonances.
We shall evaluate Eq. (\ref{eq:heme}) in some detail, but present the
general form of the final result.

The above notation is general, our model is specified by defining the
operators T$_H^{(i)}$ and T$_j$.
We assert that the action of T$_H^{(i)}$ on a nucleon leads to the
formation of a PLC which does not interact:
\begin{equation}
T_H|N\rangle=|{\rm PLC}\rangle,
\end{equation}
where only a single baryon is involved and the label $i$ is superfluous.
The PLC is a coherent sum of
three
states so that
\begin{equation}
|{\rm PLC}\rangle = \alpha |N\rangle + \beta |N^*\rangle + \gamma |N^{**}
\rangle,
\end{equation}
where $\alpha$, $\beta$ and $\gamma$ represent the elastic and inelastic
transition form factors in this model. These coefficients depend on Q$^2$;
here we assume  each has the same functional form. Thus the Q$^2$
dependences of each of the
electromagnetic form factors are the same and disappear when considering
ratios of cross sections.

Next we discuss
T$_j$. We treat this operator in lowest order so that it is the
transition matrix $\widehat T_S$ for
the soft final state interaction between the ejected wavepacket
($G_0|{\rm PLC}\rangle$) and the second proton.
This operator is defined by the condition that
\begin{equation}
\widehat T_S |{\rm PLC}\rangle =0.
\end{equation}
It is this condition that separates models with color transparency
from other models which merely include use a matrix to include
coupling to  excited states. In particular, $\widehat T_S$
is the most general 3 $\times$ 3 matrix that annihilates the PLC:
\begin{equation}
{\widehat T}_S= \left( \begin{array}{ccc}
1 & { {-\alpha + \gamma \epsilon}\over\beta} & - \epsilon \\
{ {-\alpha^* + \gamma^*\epsilon^*}\over{\beta^*}} & \mu &
{ {|\alpha|^2 - \alpha\gamma^*\epsilon^*-\mu |\beta|^2}
\over{\beta^*\gamma}} \\
-\epsilon^* &
{ {|\alpha|^2 - \alpha^*\gamma\epsilon-\mu |\beta|^2}
\over{\beta\gamma^*}} &
{ {\mu |\beta|^2 - |\alpha|^2 + 2 {\rm Re}(\alpha \gamma^*\epsilon^*)}
\over {|\gamma|^2}}
\end{array} \right).
\end{equation}
The work of Ref. \cite{FGMS} showed how to use
 data for cross section fluctuations in proton-proton diffractive
scattering to constrain
the parameters of this matrix.
We will use the parameter sets of  that work.

It is useful to present the coordinate space representations for the
operators $T_j$, $G$ and $T_H^{(1)}$.
The final-state interaction is specified by
\begin{eqnarray}
\langle N_1, {\vec R}_1; N_2, {\vec R}_2; N_3, {\vec R}_3 | & T_1 & |
m_1, {\vec R}_1; m_2, {\vec R}_2; m_3, {\vec R}_3' \rangle = \nonumber\\
&& \quad
\left({\widehat T}_S\right)_{m_1,N_1} \delta_{m_2,N_2} \delta_{m_3,N_3}
\delta^{(3)}({\vec R}_1 - {\vec R}_2)
\delta^{(3)}({\vec R}_3 - {\vec R}_3'),
\label{eq:heu}
\end{eqnarray}
where
the range of
this two-body optical potential is assumed to be exactly zero, as specified
by the above delta function.

The Green's function operator, which is diagonal in the
hadronic
 mass
eigenstate basis, is given by:
\begin{eqnarray}
\langle m_1, {\vec R}_1; N_2, {\vec R}_2; N_3, {\vec R}_3 | & G & |
m_1, {\vec R}_1'; N_2, {\vec R}_2'; N_3, {\vec R}_3' \rangle =
\nonumber \\
&&
- {{ e^{i p_{1m} | {\vec R}_1 - {\vec R}_1' | }}\over{ 4 \pi
| {\vec R}_1 - {\vec R}_1' | }}
\delta^{(3)}({\vec R}_2 - {\vec R}_2')
\delta^{(3)}({\vec R}_3 - {\vec R}_3').
\label{eq:heg}
\end{eqnarray}
Here, the quantity $p_{1m}$ is the momentum of the
$m'th$ component of the wavepacket.  We take all of the final state
nucleon wave functions as plane waves so that
energy conservation yields
\begin{equation}
|\vec p_{1m}| = \sqrt{ \left( \frac{Q^2}{2M_N} + 3M_N - \sqrt{p_2^2+M_N^2}
- \sqrt{p_3^2+M_N^2} \right)^2 - M_m^2}.
\end{equation}
In this sense, $\vec p_1 \equiv \vec p_{1N}$.
Lastly, the hard scattering operator is given by
\begin{equation}
\langle m_1, {\vec R}_1; N_2, {\vec R}_2;, N_3, {\vec R}_3 | T_H^{(1)} |
^3He \rangle = F_{m,N}(Q^2) e^{i {\vec q} \cdot {\vec R}_1 } \psi_{^3He}(
{\vec R}_1, {\vec R}_2, {\vec R}_3 ).
\label{eq:heth}
\end{equation}
Here, the quantities $F_{m,N}(Q^2)$ are the elastic and inelastic
transition form factors, $\alpha$, $\beta$ and $\gamma$.  Since only ratios
of these quantities ultimately appear, we neglect the possible dependence
on $Q^2$.  We have also introduced the position space $^3He$ wavefunction,
$\psi_{^3He}$.

We put all of the pieces together, insert
complete sets of states and arrive at the following
expression for the scattering matrix element:
\begin{eqnarray}
 {\cal M}_1(\vec p_1, \vec p_2, \vec p_3)
 & = & i \, |\vec p_1| \sum_m \left( {\widehat T}_S \right)_{N,m} F_{m,N}(Q^2)
\int (d^9R)_{cm} e^{ -i {\vec p}_1 \cdot {\vec R}_2
 -i {\vec p}_2 \cdot {\vec R}_2
 -i {\vec p}_3 \cdot {\vec R}_3 } e^{ i {\vec q} \cdot {\vec R}_1}
\nonumber \\
&& \qquad \qquad \qquad
\times {{ e^{i p_{1m} | {\vec R}_1 - {\vec R}_2 | }}\over{ 4 \pi
| {\vec R}_1 - {\vec R}_2 | }}
\psi_{^3He}({\vec R}_1, {\vec R}_2, {\vec R}_3 ),
\label{eq:hemcm}
\end{eqnarray}
where $d^9R = d^3R_1 d^3R_2 d^3R_3$ and the subscript $cm$ is to remind us
that the center of mass
of the $^3He$ nucleus, which is irrelevant for these
sorts of considerations, is factored out of the problem.
To this end, we introduce the usual Jacobi coordinates,
\begin{eqnarray}
{\vec R}_1 & = & {\vec R}_{cm} - \sqrt{1\over2} {\vec \rho} + \sqrt{
1\over6}{\vec \lambda}, \\
{\vec R}_2 & = & {\vec R}_{cm} + \sqrt{1\over2} {\vec \rho} + \sqrt{
1\over6}{\vec \lambda}, \\
{\vec R}_3 & = & {\vec R}_{cm} - \sqrt{2\over3}{\vec \lambda}.
\label{eq:jacobi}
\end{eqnarray}
Now, the above matrix element can be expressed purely in terms of the
relative coordinates, $\vec \rho$ and $\vec \lambda$.  The last piece of
information we need to specify before the calculation can proceed is the
configuration space representation of the helium wavefunction we use.  We
take the simple parameterization
\begin{equation}
\psi_{^3He}(\rho, \lambda) = \frac{\alpha^3}{\pi^{3/2}}
e^{-\alpha^2 \left( \rho^2 + \lambda^2 \right) / 2 }.
\label{eq:hewf}
\end{equation}
Here, we choose the parameter $\alpha = 1.5$ $fm^{-1}$.
Because of the delta functions which appear in
Eq.~(\ref{eq:hemcm}), the 18 dimensional integral reduces to a
simple integral over $d^3\rho d^3\lambda$, a vast improvement.  With the
above choice for the helium wavefunction, the $\lambda$ and $\rho$
integrals separate.
The resulting amplitude for producing a final state baryon
$m = N,N^*, N^{**}$ is given by
\begin{equation}
{\cal M}_1(\vec p_{1m}, \vec p_2, \vec p_3)
= \sum_{m'} \left( {\widehat T}_S \right)_{m,m'} F_{m',N}(Q^2)
{ {\alpha^3}\over{4\sqrt{2}\pi^{5/2}}} I_\lambda(\vec p_1, \vec p_2, \vec p_3)
I_\rho(\vec p_1, \vec p_2),
\label{eq:ilamirho}
\end{equation}
where
\begin{eqnarray}
I_\lambda(\vec p_1, \vec p_2, \vec p_3) & = & \int d^3\lambda e^{-\alpha^2
\lambda^2/2}
e^{i\sqrt{1/6}{\vec \lambda}\cdot \left( 2{\vec p}_3 - {\vec p}_1
-{\vec p}_2 + {\vec q} \right) } \\
I_\rho(\vec p_1, \vec p_2) & = & \int d^3\rho {1\over\rho} e^{-\alpha^2
\rho^2/2}
e^{-i \sqrt{2} p_{1m} \rho } e^{i \sqrt{1/2} {\vec \rho} \cdot
\left( {\vec q} + {\vec p}_1 + {\vec p}_2 \right) }.
\label{eq:xilamirho}
\end{eqnarray}

We can use these  expressions to explain why the effects of the
using a different t-dependence for
the different amplitudes $\alpha,\beta,\gamma$ are not very
large. Including such these ``finite range effects" would
modify the integral
$I_\rho$. The factor $1/\rho e^{-i \sqrt{2} p_{1m}\rho}$
could be replaced by, for example,
$1/\rho e^{-i \sqrt{2} p_{1m}\rho} - 1/\rho e^{-i \Lambda_m \rho}$
where $\Lambda_m$ represents the t-dependence of the relevant amplitudes.
The finite size effects are determined by the differences between the
parameters
$\gamma_m=\alpha \Lambda_m$.
Each of the $\gamma_m$ is fairly small
because the size of $^3$He is larger
than the range of the interactions in $\widehat T_S$.
The differences of small quantities are even smaller. Thus
it seems safe to neglect such differences, at least if are considering
small-t rescattering with $-t~\approx~0.15~GeV^2$.

We return to the evaluation of ${\cal M}_1$ by realizing both
$I_\rho$ and $I_\lambda$
can be done analytically and the
resulting
matrix element can be
put in closed form.  The result for the (e,e'pp) process is that
\begin{eqnarray}
{\cal M}_1(\vec p_1, \vec p_2, \vec p_3)
& = & \frac{p_1 \sqrt{\pi}}{2\alpha s} \sum_m \left(
 {\widehat T}_S
\right)_{N,m} F_{m,N}(Q^2) e^{-\frac{v^2}{2\alpha^2}}
\nonumber \\
&& \times \left[ e^{-x_+^2} \, {\rm erfc}\left( -ix_+ \right) -
e^{-x_-^2} \, {\rm erfc}\left( -ix_- \right) \right] ,
\label{eq:mclosed}
\end{eqnarray}
where $s=|\vec s|$, $v=|\vec v|$ and
\begin{eqnarray}
\vec v &=& \frac{1}{\sqrt{6}} \left( 2 {\vec p}_3 - {\vec p}_1 - {\vec p}_2
+ {\vec q} \right),
\label{eq:mclosedv}\\
\vec s &=& \frac{1}{2} \left( {\vec q} + {\vec p}_1 + {\vec p}_2 \right),
\label{eq:mcloseds}\\
x_+ &=& \frac{p_{1m}+s}{\alpha}, \label{eq:mclosedxp}\\
x_- &=& \frac{p_{1m}-s}{\alpha}, \label{eq:mclosedxm}\\
i \, {\rm erfc}(-ix) &=& i - \frac{2}{\sqrt{\pi}} \int_0^x dt \, e^{+t^2}.
\label{eq:mclosedphi}
\end{eqnarray}
The CT cross section for double scattering events
is then calculated by taking the absolute square of
this quantity, $\sigma^{(e,e'pp)} \sim |{\cal M}_1|^2$.
To see that this matrix element gives CT, we
consider the very high energy limit.  At very high energies, $x_-
\rightarrow 0$ and $x_+$ gets big.  The Gaussian factor of $e^{-x_+^2}$
kills the rapidly growing error function completely.  Then, since
${\rm erfc}(0)=1$, the matrix element reduces to some numbers times the
quantity $\sum_m \left({\widehat T}_S\right)_{N,m} F_{m,N} = 0$ and CT is
obtained.

This closed form is amusing, but does not provide much insight
and is also difficult to evaluate.
Therefore, we use Eq.~(\ref{eq:ilamirho}) and evaluate $I_\lambda$ analytically
and $I_\rho$ numerically. We need to specify the kinematics
in order to proceed.
We choose the
case where the photon hits one proton which moves quickly through the
remaining nuclear medium and interacts with the other proton which leaves
with momentum $|{\vec p}_2| = 400 \, MeV/c$, while the neutron remains a
spectator, $|{\vec p}_3| = 0$.  With these kinematics, $\vec s = \vec q$ and
$\vec v = 0$.  As noted previously, the helium bound state
parameter is taken to be $\alpha= 1.5 \, fm^{-1}$.

Now, since we are calculating a cross section which should vanish in the
limit of full transparency, we normalize our CT cross sections to the DWBA
or Glauber cross sections.
The Glauber process is defined by first using $\widehat T_S$ for the amplitude
to
produce an excited  state. Then one uses an optical potential (a diagonal
operator) to describe
the escape of the given excited state from the nucleus.
Thus the Glauber amplitude ${\cal M}_1^G(\vec p_{1m}, \vec p_2, \vec p_3)$
is given by
\begin{equation}
{\cal M}_1^G(m, \vec p_{1m}, \vec p_2, \vec p_3)
=  \left( {\widehat T}_S \right)_{m,m} F_{m,N}(Q^2)
{ {\alpha^3}\over{4\sqrt{2}\pi^{5/2}}} I_\lambda(\vec p_1, \vec p_2, \vec p_3)
I_\rho(\vec p_1, \vec p_2).
\label{eq:ilamirhog}
\end{equation}
This so-called distorted wave Born approximation to the amplitude leads to
DWBA cross sections  labelled as $\sigma_{Glauber}$
Thus, for the nucleon, the cross section ratio starting
out at unity, where we know the DWBA works well, and decreasing to zero at
high momentum transfers.  The rate of decrease depends on the initial
wavepacket or PLC.

The motivation for the calculation of the double scattering events is that
the Glauber treatment of the single scattering events, $(e,e'p)$ events,
is already close to the plane wave values.  But we have argued that
examining  processes where the cross section vanishes in the CT
limit and the $Q^2$ variation is rapid at moderate values of $Q^2$
increases the sensitivity to CT effects.
Figures~1~and~2 (d) show that we
can further
increase the $Q^2$ variation by including these relatively
slowly varying single scattering events by considering the ratio
\begin{equation}
(T_{el}/T)_{CT/G}
\equiv \frac{ \sigma^{CT}_{(e,e'pp)} / \sigma^{Glauber}_{(e,e'pp)}}
{\sigma^{CT}_{(e,e'p)} / \sigma^{Glauber}_{(e,e'p)}}.
\end{equation}
This is the ratio of $T_{el}/T$ for CT to Glauber calculations
shown in Figs.~1~and~2~(d). This expression is generalized for the case of
resonance production by replacing the ``$pp$" by N$^*p$ or N$^{**}p$.

In order to calculate the cross section for single scattering events, we
must go back to the original amplitude and calculate the Born term.  Thus,
we write the full amplitude, to first order in the interaction, as a
sum of two terms
\begin{equation}
{\cal M} (\vec p_1,\vec p_2, \vec p_3) =
{\cal M}_0 (\vec p_1, \vec p_2, \vec p_3) +
{\cal M}_1 (\vec p_1, \vec p_2, \vec p_3),
\end{equation}
where ${\cal M}_1 (\vec p_1, \vec p_2, \vec p_3)$ is given, generally, by
Eq.~(\ref{eq:mclosed}).  It is easy to calculate the Born term.  The result
is
\begin{equation}
{\cal M}_0(\vec p_1,\vec p_2,\vec p_3) = \frac{8 \pi^{3/2}}{\alpha^3}
e^{-\left( w^2 + v^2 \right)/2\alpha^2} F_{NN}(Q^2),
\end{equation}
where $v=|\vec v|$ in Eq.~(\ref{eq:mclosedv}), $w=|\vec w|$ and
\begin{equation}
\vec w = \frac{1}{\sqrt{2}} \left( \vec p_1 - \vec p_2 - \vec q \right).
\end{equation}
Now, the $(e,e'p)$ cross section has different kinematics than the
$(e,e'pp)$ one.  In particular, for single nucleon knockout, we imagine
that the detected proton is carrying all of the photon's three-momentum,
which leaves very little for the other two nucleons.  Thus, we imagine that
$\vec p_1 = \vec q$ and $\vec p_2 = \vec p_3 =0$.  In these kinematics
$\vec w=\vec v=0$.  The relevant ratio we
want to consider then, is
\begin{equation}
(T_{el}/T)_{CT/G}= \left| \frac{
{\cal M}^{CT/G}_1(\vec p_1, |\vec p_2|=400\, MeV,0)  }{
{\cal M}^{CT/G}(\vec p_1,0,0)  } \right|^2,
\end{equation}
where the above notation means
\begin{equation}
{\cal M}^{CT/G} (\vec p_1, \vec p_2, \vec p_3) \equiv
\frac{ {\cal M}^{CT}(\vec p_1, \vec p_2, \vec p_3) }{
{\cal M}^{Glauber}(\vec p_1, \vec p_2, \vec p_3)}.
\end{equation}

We present results for
the quasi-elastic production of nucleons, $N^*$'s and $N^{**}$'s for two
sets of masses of the excited states, in the form of ratios
$\Sigma_{2/G;1/G}$.
The generalization of
Eq.~(\ref{eq:heme}) to the quasi-elastic production of nucleon isobars
is straightforward.
In Figure~4
we take
${M_{N^*} = 1.44 \, GeV}$ and ${M_{N^{**}} = 1.80 \, GeV}$ while in
Figure~5 we take
${M_{N^*} = 1.80 \, GeV}$ and ${M_{N^{**}} = 3.0 \, GeV}$. The values of
these masses control the rate of PLC expansion and therefore the energy
dependence of the computed ratios of cross sections.
The first set, with the lower values of masses, leads to
a slow rate of expansion and larger CT effects, while the second set
has a quicker rate of expansion.
The parameters of the matrix $\widehat T_S$ are specified in the figures
and in Ref.~\cite{FGMS}.

Fig. 4 shows that $\Sigma_{2/G;1/G}$ decreases
by factors between
2 and 10, as Q$^2$ is varied from about 1.5 to 7 GeV$^2/$c$^2$.
These factors are reduced somewhat by increasing the values of the
excited state masses, an expected effect.
The use of the parameter set shown on the lower right
along with of higher values of
the masses leads to a result with little variation with Q$^2$.
This is the only such case.
We also
note that the $(e,e'p)$ Glauber cross section for the nucleon (which is
essentially the same as the CT cross section at low energies), using the
above equations, remains constant at approximately $0.68$.  This number is
just controlled by the value of the $pp$ cross section $\sigma = 40 \, mb$.

The advantage of using double scattering events, as described here,
 to explore the effects of
CT is clear upon looking at the figures.  We see that there are
significant effects, as large as an order of magnitude,
 for values of $Q^2$ as
low as $Q^2$ between $\approx 6 $ and 8 GeV$^2$/c$^2$.

\section{Summary and Conclusions}

We have shown that the effects of color transparency can be investigated at
intermediate values of $Q^2 \geq 5-6 ~GeV^2$ by detecting a final quasielastic
proton and
another with momentum about 300-400~$ MeV/c$. We have used oversimplified wave
functions,
which do not allow us to investigate the effects of
the nuclear correlations .
This could be corrected in future studies. Most important
would be detailed experimental study of this reaction which is feasible at
CEBAF.

\vfill\eject
{\large{\bf Figure Captions}}
\begin {itemize}

\item [Figure 1.] The dependence of nuclear transparencies to $Q^2$ for
$^4He$ target. {\bf (a)} - transparencies for $(e,e'p)$ processes,
solid -- Glauber  approximation, dashed -- color transparency
approximation. {\bf (b)} -- Transparency $T_{el}$ as defined
in Eq.~(\ref{eq:T-el}), for $(e,e'pp)$ reactions: notations same as in (a).
{\bf (c)} -- ratio of transparencies ($T_{el}\over T$) defined in
Eq.~(\ref{eq:T_r}): notations same as in (a).
{\bf (d)} -- Ratios of $T$, $T_{el}$ and $T_{el}/T$ at the case of
color transparencies to the corresponding quantities at  Glauber
approximations; Solid line - ${ T^{CT}\over T^{GA}}$, dashed line
- ${ T_{el}^{CT}\over T_{el}^{GA}}$ and dotted line -
${ (T_{el}/T)^{CT}\over (T_{el}/T)^{GA}}$.

\vspace {0.5cm}

\item [Figure 2.] Same as in Fig.1, for $^{12}C$.

\vspace {0.5cm}

\item [Figure 3.] The $t$ dependence of nuclear transparencies -
 (${\sigma(e,e'pp)^{CT}\over\sigma(e,e'pp)^{GA}}$) for $^4He$ at
$(e,e'pp)$ reaction at $Q^2 = 4 GeV^2$ (solid line) and $Q^2 = 10 GeV^2$
(dashed line), provided that slope of $pp$ cross section is changed
according to Eq.~(\ref{eq:sigma-pp}).

\item [Figure 4.]
Cross sections ratios $\Sigma_{2/G;1/G}$ in three-state model for $^3He$.
Solid: quasi-elastic
proton production. Dashed: quasielastic N$^*$ (1.4 GeV) production. Dotted:
quasielastic N$^{**}$ (1.8 GeV) production. The parameters $\alpha,\beta,
\gamma$ and $\epsilon$ define the baryon-nucleon interaction, see
Ref.~\cite{FGMS}.
\vspace{0.5cm}

\item [Figure 5.]
The same as in Fig. 4 but the masses are now (N$^*$,N$^{**}$)=(1.8, 3.0) GeV.
\vspace{0.5cm}

\end{itemize}
\end{document}